\documentclass[%
 reprint,
 amsmath,amssymb,
 aps,
]{revtex4-2}

\usepackage{mathtools}
\usepackage{amssymb}
\usepackage{bm}
\usepackage{braket}
\usepackage{natbib}
\usepackage{graphicx}
\usepackage{dcolumn}
\usepackage{bm}

\begin{document}

\preprint{APS/123-QED}

\title{On the quantization of the extremal Reissner-Nordstrom black hole}

\author{Christian Corda}
\email{cordac.galilei@gmail.com}
\affiliation{%
International Institute for Applicable Mathematics and Information Sciences
(IIAMIS), B.M. Birla Science Centre, Adarsh Nagar, Hyderabad -- 500 463, India
}%
\author{Fabiano Feleppa}
\email{feleppa.fabiano@gmail.com}
\affiliation{
Institute for Theoretical Physics, Utrecht University,
Princetonplein 5, 3584 CC Utrecht, The Netherlands
}%
\author{Fabrizio Tamburini}
\email{fabrizio.tamburini@gmail.com}
\affiliation{ZKM -- Zentrum f\"ur Kunst und Medientechnologie, Lorentzstr. 19, D-76135, Karlsruhe, Germany}%


\begin{abstract}
Following Rosen's quantization rules, two of the Authors (CC and FF)
recently described the Schwarzschild black hole (BH) formed after
the gravitational collapse of a pressureless ``star of dust'' in
terms of a ``gravitational hydrogen atom''. Here we generalize this
approach to the gravitational collapse of a charged object, namely,
to the geometry of a Reissner-Nordstrom BH (RNBH) and calculate the
gravitational potential, the Schr\"odinger equation and the exact solutions
of the energy levels of the gravitational collapse. By using the concept
of \emph{BH effective state}, previously introduced by one of us (CC),
we describe the quantum gravitational potential, the mass spectrum
and the energy spectrum for the extremal RNBH. 
The area spectrum derived from the mass spectrum finds agreement with a previous result by Bekenstein.
The stability of these solutions, described with the Majorana approach
to the Archaic Universe scenario, show the existence of oscillatory
regimes or exponential damping for the evolution of a small perturbation
from a stable state.
\end{abstract}

\maketitle

\section{Introduction}

Black holes are at all effects theoretical and conceptual laboratories where
one discusses, tests and tries to understand the fundamental problems
and potential contradictions that arise in the various attempts made
to unify Einstein's general theory of relativity with quantum mechanics.
In a previous paper \cite{key-1}, two of us (CC and FF) suggested a new model
of quantum BH based on a mathematical analogy with the hydrogen
atom obtained by using the same quantization approach proposed in
1993 by the historical collaborator of Einstein, N. Rosen \cite{key-2}. It is well-known that the canonical quantization of general relativity leads to the Wheeler-DeWitt equation introducing the so-called Superspace, an infinite-dimensional space of all possible 3-metrics; Rosen, instead, preferred to start his work from the classical cosmological equations using a simplified quantization procedure. In this view, BHs should behave as regular quantum systems with a
discrete energy spectrum, finding similarities with Bekenstein's results
\cite{key-3}. The quantum properties of BHs are ruled by a Schr\"odinger
equation with a wave function representation that could play an important
role in the solution of the BH information paradox too \cite{key-4}.
This approach, even if it presents some mathematical analogies that
resemble the behavior of the hydrogen atom, is different from that adopted
in \cite{key-5}, where the evolution of the BH microstates are described
by a perturbative quantum field theory defined in a Schwarzschild
metric background and a cut-off is required to replace the transverse
coordinates by a lattice.

Rosen’s approach has also been recently applied to a cosmological framework \cite{Feleppa, Feoli} and to the famous Hartle-Hawking initial state \cite{Feleppa2}.

In this paper, we extend the results of \cite{key-1} to the class
of extremal RNBHs, starting from the gravitational collapse of a set
of dust particles with electrostatic charge. Also in this case, the
study of the collapse of a simple set of dust particles can be a first
and valid approach to face the problem as it presents the advantage
of finding and highlighting some fundamental properties (or
analogies) that can be found also in other models characterized by, e.g.,
more articulated descriptions in terms of quantum fields that require
a more complicated interpretation. In any case, this method, even if
simplified, has a good validity as, up to now, there is no unique
and valid theory of quantum gravity that suggests a specific model
to use.

In Rosen's approach to the Schwarzschild solution, the ground state
can be interpreted in terms of phenomenology at Planck scales, with
the caveat that the Planck mass found in the ground state can be interpreted
in terms of energy fluctuations, graviton exchange equivalent to wormhole
connections avoiding the presence of a singularity in a flat background
(see \cite{key-6} for details). This means that when the BH evaporates it leaves a flat
spacetime where quantum fluctuations occur, in agreement with the
standard scenario.

Thanks to the simplicity of the model used here, within Rosen's framework, we find the gravitational potential, the Schr\"odinger equation and
the solution for the energy levels' collapse that lead to a RNBH by
focussing our attention to the extremal case. Also here we find that
this simple model obeys the rules of a hydrogen atom and the extremal
RNBH mass spectrum is found as an exact solution. From the mass spectrum,
the area spectrum is also obtained. What is really interesting is
that, in any case, one finds that the energy levels, the evolution
of the quantum microstates and the evolution of the perturbations
of any energy level found obey a mathematical structure similar to
that of the hydrogen atom as it was in the early days of quantum mechanics; this is essentially due to the spherical symmetry of the system and
the solutions may vary from model to model depending on the mathematical
shape of the function that describes the fields and acts as potential
in the Schr\"odinger equation. The advantage of analyzing such an oversimplified
model is that it can give some interesting hints to the understanding
of the phenomenology at the Planck scale. 

\section{Application of Rosen's quantization approach to the gravitational
collapse of a charged BH}

Gravitational collapse is one of the most important problems in
general relativity. In particular, a complete description of the gravitational
collapse of a charged perfect fluid in the class of Friedmann universe models can
be found in \cite{key-7}. To describe the interior of the collapsing
star one can use the well-known Friedmann-Lemaitre-Robertson-Walker
(FLRW) line element that can be written, in Planck units 
($G=c=k_{B}=\hbar=1/4\pi\epsilon_{0}=1$),
as in \cite{key-8,key-9}:
\begin{equation}
ds^{2}=d\tau^{2}-a^{2}(t)[d\chi^{2}+\sin^{2}\chi(d\theta^{2}+\sin^{2}\theta d\phi^{2})],
\end{equation}
where $a(t)$ is the scale factor. Notice that, by setting the value
for $\sin^{2}\chi$, one chooses the case of a spacetime with positive
curvature, which corresponds to a gas sphere whose dynamics starts
at rest at a given finite radius. On the other hand, the external
geometry is described by the Reissner-Nordstrom line-element \cite{key-7}
\begin{equation}
ds^{2}=Zdt^{2}-\frac{1}{Z}dr^{2}-r^{2}(d\theta^{2}+\sin^{2}\theta d\phi^{2})),
\end{equation}
where
\begin{equation}
Z(r)=1-\frac{2M}{r}+\frac{Q^{2}}{r^{2}}.\label{eq:3}
\end{equation}
In Eq. ($\ref{eq:3}$), the quantity $M$ represents the total mass of the
collapsing star and the parameter $Q$ is its electrostatic charge.
The internal homogeneity and isotropy of the FLRW line-element
break up at the star's surface at a certain given radius $\chi=\chi_{0}$.
Thus, one has to consider a range of values for the parameter $\chi$ defined
in the interval $0\le\chi\le\chi_{0}$, with $\chi_{0}<\frac{\pi}{2}$
during the collapse, with the requirement that the interior
FLRW geometry matches the exterior Reissner-Nordstrom geometry. Such
a matching is given by setting \cite{key-7}
\begin{equation}
r=a\sin\chi
\label{eq:4}
\end{equation}
and 
\begin{equation}
M=\frac{a}{2}\sin\chi+\frac{Q^{2}}{2a\sin\chi}+\frac{a\dot{a}}{2}\sin\chi^{3}-\frac{a}{2}\sin\chi\cos\chi^{2}.\label{eq:5}
\end{equation}
Eqs. (\ref{eq:4}) and (\ref{eq:5}) hold only on the matching surface, i.e., they provide the necessary and sufficient conditions for the smooth matching of the two regions over such a surface \cite{key-7}. Time derivation in Eq. (5) is made with respect to the proper time at the star surface \cite{key-8}, that will also be the time considered hereafter. 

In the following we will apply Rosen's quantization approach \cite{key-2}, who described a closed, homogeneous and isotropic universe,
to that of a collapsing star leading to the formation of a RNBH, and discuss in detail the results.

By solving the Einstein-Maxwell field equations, one can find a system analogous to the Friedmann equations with the
additional terms derived from the electromagnetic stress-energy tensor \cite{key-10}:
\begin{eqnarray}
3\left(\frac{1+\dot{a}^{2}}{a^{2}}\right)&=&4\pi\frac{E_{0}^{2}}{a^{4}}+8\pi\rho,\label{eq:6}
\\ 
-2\frac{\ddot{a}}{a}-\left(\frac{1+\dot{a}^{2}}{a^{2}}\right)&=&-4\pi\frac{E_{0}^{2}}{a^{4}}+8\pi\rho,
\\
-2\frac{\ddot{a}}{a}-\left(\frac{1+\dot{a}^{2}}{a^{2}}\right)&=&4\pi\frac{E_{0}^{2}}{a^{4}},\label{eq:8}
\end{eqnarray}
having set the cosmological constant term $\Lambda=0$ and $E_{0}^{2}\equiv\frac{Q}{\left(a\sin\chi\right)^{2}}$.

As it turns out, we can deal with these equations by using a similar
procedure as in the standard case \cite{key-7}. Eq. ($\ref{eq:6}$)
becomes equivalent to 
\begin{equation}
\frac{a}{2}+\frac{a\dot{a}^{2}}{2}-\frac{2\pi}{3}\frac{E_{0}^{2}}{a}=\frac{4\pi}{3}\rho a^{3}.\label{eq:9}
\end{equation}
After taking the time derivative of both sides and using  Eq. (\ref{eq:8}),
we find that
\begin{equation}
\rho=\frac{3M_{*}}{4\pi a^{3}}+\frac{E_{0}}{a^{4}},\label{eq:10}
\end{equation}
where $M_{*}$ is an integration constant. In the absence of electromagnetic
field (i.e., for $E_{0}=0$) we find, as in \cite{key-1}, that
\begin{equation}
\rho=\frac{3a_{0}}{8\pi a^{3}}.
\end{equation}
Hence, we can deduce that the integration constant is
\begin{equation}
M_{*}=\frac{a_{0}}{2}.
\end{equation}
Finally, we obtain the expression of the density as 
\begin{equation}
\rho=\frac{3a_{0}}{8\pi a^{3}}+\frac{E_{0}}{a^{4}}.
\end{equation}
Using Eq. ($\ref{eq:9}$) and Eq. ($\ref{eq:10}$), and multiplying by $M/2$, one obtains 
\begin{equation}
\frac{1}{2}M\dot{a}^{2}-\left(\frac{2M\pi E_{0}^{2}}{a^{2}}+\frac{Ma_{0}}{2a}\right)=-\frac{M}{2},
\end{equation}
which can be interpreted as an energy equation for a particle in one-dimensional
motion having coordinate $a$:
\begin{equation}
E=T+V,
\end{equation}
where the kinetic energy is 
\begin{equation}
T=\frac{1}{2}M\dot{a}^{2},
\end{equation}
and the potential energy is 
\begin{equation}
V(a)=-\left(\frac{2M\pi E_{0}^{2}}{a^{2}}+\frac{Ma_{0}}{2a}\right).\label{eq:17}
\end{equation}
Thus, the total energy is 
\begin{equation}
E=-\frac{M}{2}.\label{eq:18}
\end{equation}
From the Friedmann-like equation already discussed one obtains the
equation of motion of the particle with momentum given by 
\begin{equation}
P=M\dot{a},
\end{equation} 
and associated Hamiltonian 
\begin{equation}\label{eq:20}
\mathcal{H}=\frac{P^{2}}{2M}+V.
\end{equation}

Up to now, we faced the problem of the gravitational collapse from the classical point of view only.
In order to discuss it also from the quantum point of view, we then need
to define a wave-function 
\begin{equation}
\Psi\equiv\Psi(a,\tau).\label{eq: 21}
\end{equation}
Thus, in correspondence of the classical equation (\ref{eq:20}),
one gets the traditional Schr\"odinger equation 
\begin{equation}
i\frac{\partial\Psi}{\partial t}=-\frac{1}{2M}\frac{\partial^{2}\Psi}{\partial a^{2}}+V\Psi.\label{eq:22}
\end{equation}
For a stationary state with energy $E$ one obtains
\begin{equation}
\Psi=\Psi(a)e^{-iE\tau},
\end{equation}
and equation ($\ref{eq:22}$) becomes 
\begin{equation}
-\frac{1}{2M}\frac{\partial^{2}\Psi}{\partial a^{2}}+V\Psi=E\Psi.\label{eq:24}
\end{equation}
Besides, by considering Eq. ($\ref{eq:17}$), equation ($\ref{eq:24}$)
can be written as 
\begin{equation}
-\frac{1}{2M}\frac{\partial^{2}\Psi}{\partial a^{2}}+\left(-\frac{2\pi ME_{0}^{2}}{a^{2}}-\frac{Ma_{0}}{2a}\right)\Psi=E\Psi.\label{eq:25}
\end{equation}
Now, by setting 
\begin{equation}
A=-2\pi ME_{0}^{2},\hspace{1cm}B=\frac{Ma_{0}}{2},
\end{equation}
we can write
\begin{equation}
-\frac{1}{2M}\frac{\partial^{2}\Psi}{\partial a^{2}}+\left(\frac{A}{a^{2}}-\frac{B}{a}\right)\Psi=E\Psi.
\end{equation}
This potential has been deeply studied in \cite{key-11},
where the authors found the energy spectrum. In our case we have
\begin{eqnarray}
E_{n}&=&-\frac{2MB^{2}}{\left(2n-1+\sqrt{1+8A}\right)^{2}} \nonumber
\\
 &=& -\frac{M^{3}a_{0}^{2}}{2\left(2n-1+\sqrt{1-16\pi ME_{0}^{2}}\right)^{2}},\label{eq:28}
\end{eqnarray}
where $n$ is the principal quantum number. It should be noted that,
when $E_{0}=0$, we recover the result which has been found in \cite{key-1}:
\begin{equation}
E_{n}=-\frac{M^{3}a_{0}^{2}}{8n^{2}}.\label{eq:29}
\end{equation}
Following Rosen's approach, one then inserts Eq. ($\ref{eq:18}$) into
Eq. ($\ref{eq:28}$), obtaining 
\begin{equation}
-\frac{2M^{3}a_{0}^{2}}{4(2n-1+\sqrt{1-M16\pi E_{0}^{2}})^{2}}=-\frac{M}{2}.\label{eq: 30}
\end{equation}
The last equation admits only two acceptable solutions, namely
\begin{align}
M_{1n}=&-\frac{\sqrt{4a_{0}^{2}-8a_{0}n\beta+4a_{0}\beta+\beta^{2}}}{2a_{0}^{2}}\nonumber \\
&+\frac{4a_{0}n-2a_{0}-\beta}{2a_{0}^{2}},\label{eq: 31}
\end{align}
and 
\begin{align}
M_{2n}=&\frac{\sqrt{4a_{0}^{2}-8a_{0}n\beta+4a_{0}\beta+\beta^{2}}}{2a_{0}^{2}}\nonumber \\
&+\frac{4a_{0}n-2a_{0}-\beta}{2a_{0}^{2}},\label{eq:32}
\end{align}
where $\beta=16\pi E_{0}^{2}$. 
If $E_{0}=0$ (i.e., $\beta=0$), then one obtains 
\begin{equation}
M_{1n}=\frac{4a_{0}(n-1)}{2a_{0}^{2}}=\frac{2(n-1)}{a_{0}}, \label{eq:33}
\end{equation}
and
\begin{equation}
M_{2n}=\frac{4a_{0}n}{2a_{0}^{2}}=\frac{2n}{a_{0}}.\label{eq:34}
\end{equation}
The second solution, $M_{2n}$, recovers the previous result presented
in \cite{key-1}, where the ground state of the collapsing star is
found by setting $n=1$. Now, from Eq. ($\ref{eq:33}$),
for $n=1$, we obtain a mass spectrum equal to zero, which is
unacceptable from a physical point of view. 
Therefore, one has to consider as solution only the value $M_{2n}$ that we 
will write from now as $M_{n}$ for the sake of simplicity.
On the other hand, by using Eq. ($\ref{eq:18}$), one finds the energy
levels of the collapsing star as 
\begin{align}
E_{n}=&-\frac{\sqrt{a_{0}^{2}-2a_{0}n\beta+a_{0}\beta+(\beta/2)^{2}}}{2a_{0}^{2}}\nonumber \\
&+\frac{2a_{0}n-a_{0}-\beta/2}{2a_{0}^{2}}.\label{eq:35}
\end{align}
Eq. ($\ref{eq:32}$) represents the mass spectrum of the collapsing
star, while Eq. ($\ref{eq:35}$) represents its energy spectrum, where
the gravitational energy, given by Eq. (\ref{eq:17}), is included.

\section{Energy spectrum and ground state for the extremal RNBH}

Let us consider the case of a completely collapsed star, i.e., a RNBH.
Setting $\chi=\pi/2$, and evaluating Eqs. ($\ref{eq:4}$)
and ($\ref{eq:5}$) at $\tau=0$, we obtain 
\begin{equation}
r=a\;\;r_{i}=a_{0}=M+\sqrt{M^{2}-Q^{2}},\label{eq: 36}
\end{equation}
having set the initial velocity of the collapse equal to zero.
Moreover, now we also write $E_{0}=\frac{Q}{a_{0}^{2}}$. For the sake
of simplicity, here we will discuss the extremal case, that means $M=Q.$
Thus, Eq. (36) becomes 
\begin{equation}
r=a\;\;r_{i}=a_{0}=M,\label{eq: 37}
\end{equation}
with $E_{0}=1/M$. Now it is possible to find the potential energy of an extremal
RNBH, its Schr\"odinger equation, the inert mass spectrum and the energy spectrum. 

The potential energy is found from
Eqs. ($\ref{eq:17}$), ($\ref{eq:25}$) and ($\ref{eq:28}$):
\begin{equation}
V(r)=-\left(\frac{2\pi}{Mr^{2}}+\frac{M^{2}}{2r}\right).\label{eq: 38}
\end{equation}
The Schr\"odinger equation is
\begin{equation}
-\frac{1}{2M}\frac{\partial^{2}\Psi}{\partial r^{2}}-\left(\frac{2\pi}{Mr^{2}}+\frac{M^{2}}{2r}\right)\Psi=E\Psi,\label{eq: 39}
\end{equation}
and the energy spectrum is
\begin{equation}
E_{n}=-\frac{M^{5}}{2\left(2n-1+\sqrt{1-\frac{16\pi}{M}}\right)^{2}}.\label{eq:  40}
\end{equation}
The mass spectrum of an extremal RNBH can be obtained from the energy spectrum
by using  Eq. (\ref{eq:18}). One then gets 
\begin{equation}
M_{n}^{2}=2n-1+\sqrt{1-\frac{16\pi}{M_{n}}}.
\label{eq:41}
\end{equation}
For large values of the principal quantum number $n$ of the extremal RNBH, one gets an approximated solution of Eq. (\ref{eq:41}), that
is 
\begin{equation}
M_{n}=\sqrt{2n}.\label{eq: 42}
\end{equation}
Thus the energy levels of the extremal RNBH are
\begin{equation}
E_{n}=-\frac{\sqrt{2n}}{2}.\label{eq: 43}
\end{equation}
Actually, a final further correction is needed. In fact, due to the absorptions of external
particles, the extremal RNBH mass changes during the jumps from one
energy level to another. The RNBH mass increases for energy absorptions. Therefore, one must also include this
dynamical behavior to properly describe the properties of the extremal RNBH. 

To do that, one can introduce the \emph{RNBH effective state} as in \cite{key-12}.
Let us consider the emission of Hawking quanta or the absorption of external particles. 
Starting from the seminal work by Parikh and Wilczek \cite{key-13}, 
one of us (CC) introduced the concept of 
\emph{BH effective temperature}, \emph{effective mass} and \emph{effective charge} \cite{key-14}. In the case of an extremal RNBH the mass equals the electrostatic charge; to describe it, the use of the BH \emph{effective mass} concept is sufficient. More precisely, if one considers the case of absorptions,  where $M$ is the initial extremal RNBH mass before the absorption and $M+\omega$ is the final extremal RNBH mass after the absorption of an external particle having mass-energy $\omega$,  the extremal RNBH \emph{effective mass} can be introduced as in \cite{key-12},
viz., 
\begin{equation}
M_{E}(\omega)\equiv M+\frac{\omega}{2}.
\label{eq: 44}
\end{equation}
The effective mass is defined as an averaged quantity
and actually represents the average of the initial and final BH masses \cite{key-1,key-12,key-14}
before/after the absorption.
In the present case, this averaged quantity represents the extremal
RNBH mass \emph{during} the expansion. Therefore, in order to take the dynamical behavior of an extremal RNBH into
account, in the case of increasing mass, one must replace the extremal
RNBH mass $M$ with the extremal RNBH effective mass, obtaining a potential 
\begin{equation}
V(r)=-\left(\frac{2\pi}{M_{E}r^{2}}+\frac{M_{E}^{2}}{2r}\right).\label{eq: 46}
\end{equation}
The RNBH Schr\"odinger equation becomes
\begin{equation}
-\frac{1}{2M_{E}}\frac{\partial^{2}\Psi}{\partial r^{2}}-\left(\frac{2\pi}{M_{E}r^{2}}+\frac{M_{E}^{2}}{2r}\right)\Psi=E\Psi.\label{eq: 47}
\end{equation}
The introduction of the extremal RNBH effective mass in the dynamical
equations can be rigorously justified by using Hawking's periodicity
argument \cite{key-15} (see \cite{key-1,key-16} for a deeper 
insight in the mathematical approach). 

Now, from the quantum point of view, we want to obtain the energy
eigenvalues describing absorptions phenomena that start just
after the formation of 
the extremal RNBH, 
i.e., from the ideal case 
of a RNBH having null mass. This is obtained by replacing 
$M\rightarrow0$ and $\omega\rightarrow M$ in Eq.
(\ref{eq: 44}), obtaining
\begin{equation}
M_{E}\equiv\frac{M}{2}.\label{eq: 48}
\end{equation}
This allows us to write down the final equations for the extremal RNBH
mass and energy spectra as 
\begin{equation}
M_{n}=2\sqrt{2n}\label{eq: 49}
\end{equation}
and 
\begin{equation}
E_{n}=-\frac{\sqrt{2n}}{2},\label{eq: 50}
\end{equation}
respectively.
Now, one recalls that the relationship between area, mass and charge
of the RNBH is \cite{key-3}
\begin{equation}
A=4\pi\left(M+\sqrt{M^{2}-Q^{2}}\right)^{2},
\label{eq: Bekenstein}
\end{equation}
that becomes, for the extremal RNBH, 
\begin{equation}
A=4\pi M^{2}.
\label{eq: 52}
\end{equation}

Consider now an extremal RNBH which is excited at the level $n$; if one assumes that a neighboring particle is captured
by the extremal RNBH causing a transition from the state with $n$
to the state with $n+1$, then the variation of the extremal RNBH
area becomes
\begin{equation}
\Delta A_{n\rightarrow n+1}\equiv A_{n+1}-A_{n}
\label{eq: absorbed}
\end{equation}
and, by combining Eqs. (\ref{eq: 49}) and (\ref{eq: 52}), one immediately
finds the quantum of area to be
\begin{equation}
\Delta A_{n\rightarrow n+1}=4\pi\left(M_{n+1}^{2}-M_{n}^{2}\right)=32\pi.\label{eq: 54}
\end{equation}

A similar case was analysed by Bekenstein
\cite{key-3}, who considered approximately constant the RNBH mass
during the transition, obtaining 
\begin{equation}
\Delta A=8\pi M\epsilon,
\label{eq: area quantum Bekenstein}
\end{equation}
where the quantity $\epsilon$ is the total energy of the absorbed particle. In
Bekenstein's approximation,
\begin{equation}
M = M_{n}\simeq M_{n+1}\Rightarrow2\sqrt{2n}\simeq2\sqrt{2(n+1)},\label{eq: 56}
\end{equation}
while, from Eq. (\ref{eq: 50}), one gets 
\begin{equation}
\epsilon=\left|-\frac{\sqrt{2(n+1)}}{2} + \frac{\sqrt{2n}}{2}\right|.\label{eq: 57}
\end{equation}
Thus, combining Eqs. (\ref{eq: area quantum Bekenstein}), (\ref{eq: 56})
and (\ref{eq: 57}), one obtains 
\begin{equation}
\Delta A\simeq16\pi.\label{eq: 58}
\end{equation}
This shows that our result describing the quantum of area of the extremal RNBH is consistent
with the result found by Bekenstein \cite{key-3}. 

\section{Stability of the solutions with respect to a perturbation}

The mathematical analogy of gravitational collapse leading to
an extremal RNBH with that of an hydrogen atom, presents additional
advantages that can characterize the stability of the solutions, already
found with respect to an external perturbation.
This is obtained starting from Rosen's approach to cosmology and from 
that in \cite{key-17}. The Schr\"odinger equation ruling the gravitational collapse of 
an extremal RNBH, as in the Archaic universe scenario \cite{key-18}, 
can admit the existence of oscillatory solutions induced by the perturbation
or an exponential suppression of the perturbation.
This behavior is observed also in universes represented by a dust model with cosmological
constant expressed in terms of Bessel functions. 

Consider, after simple algebra, the general mathematical form of Eq. (\ref{eq: 47}).
The generic Schr\"odinger equation for the collapse of a RNBH becomes
\begin{equation}
\psi^{\prime\prime}+P\ \psi=0,
\end{equation}
where $'$ represents the derivative with respect to the parameter
$r$, and the function $P$ is defined as
\begin{equation}
P=\frac{4\pi}{Mr^{2}}+\frac{M^{3}}{r}-\frac{M}{2}.
\end{equation}
We can study the evolution of the solutions of
Eq. (\ref{eq: 47}) by adopting the approach by Ettore Majorana in
his unpublished notes \cite{key-19} for the study of the stability
and scattering of the hydrogen-like atom as in \cite{key-18}. Thus one can write
\begin{equation}
\psi=u\exp{\left\{ i\int\frac{k_{1}}{u^{2}}dx\right\} }.
\end{equation}
The equation of motion becomes 
\begin{equation}
u^{\prime\prime}-\frac{k_{1}}{u^{2}}+P\ u=0,
\end{equation}
with the general solution 
\begin{equation}
\psi=u_{1}[A^{*}\exp(I)+B^{*}\exp(-I)],
\end{equation}
where
\begin{equation}
I \equiv i\int\frac{1}{u_{1}^{2}}dx,
\end{equation}
and $A^{*}$ and $B^{*}$ are two coefficients. 

Let us now suppose that the perturbation is small and slowly varying,
namely, $|P'/P|\ll1$. Then we can set, without loss of generality,
$u=P^{-1/4}$ and $P>0$, that is obtained in the following interval:
\begin{equation}
\frac{2M^{2}-\sqrt{M^{4}+\frac{16\pi}{M}}}{4}<r<\frac{2M^{2}+\sqrt{M^{4}+\frac{16\pi}{M}}}{4}
\end{equation}
After having set the values of $A^{*}$ and $B^{*}$ from the initial conditions assumed in the gravitational collapse, and considering 
\begin{eqnarray}
P' & = & -\frac{8\pi M^{4}}{r^{3}}-\frac{M^{3}}{r^{2}}\\
P'' & = & \frac{24\pi}{r^{4}}+\frac{2M}{r^{3}}, 
\end{eqnarray}
we clearly find oscillatory solutions \cite{key-19}. By defining
\begin{equation}
I(P) \equiv \int\sqrt{P}\left(1-\frac{P'^{2}}{32P^{3}}\right)dr,
\end{equation}
these solutions can be written as
\begin{align}
\psi=&\frac{1}{\sqrt[4]{P}}\left(1+\frac{PP''-5/4P'^{2}}{16P^{3}}\right)\nonumber \\
&\times \left\{ \left.\begin{array}{c}
\sin\\
\cos
\end{array}\right.\left[-\frac{P'}{8P^{3/2}}+I(P)\right]\right\} ,
\end{align}
which means that a small and slow (in $r$) perturbation induces an
oscillatory regime in the RNBH state. The coordinate $r$ should be positive, and so we must have 
$r>\sqrt[4]{\frac{4\pi}{M}}$. This quantity
is related to the turnaround point $r_{0}=\frac{M}{2}$ where the
infall velocity slows all the way to zero between the two horizons
of the extremal RNBH and the oscillation may occur when the coordinate
takes values $r>\sqrt[4]{4\pi}$. 

Another type of solution is found when $P<0$. By setting $P_{1}=-P>0$,
Majorana found the following solution for a small perturbation in
$r$:
\begin{align}
\psi=&\frac{1}{\sqrt[4]{P_{1}}}\left(1-\frac{P_{1}P_{1}''-5/4P_{1}'^{2}}{16P_{1}^{3}}\right)\nonumber \\
&\times \exp\left\{\pm\frac{P_{1}'}{8P_{1}^{3/2}}+I(P_1)\right\}
\end{align}
In this case the perturbation is suddenly exponentially damped, confirming
the stability of the solutions found so far.

\section{Conclusion remarks}
In this brief paper we discussed the properties and behavior of the gravitational collapse
of a charged object forming a Riessner-Nordstr\"om black hole
down to the quantum level, by calculating
the gravitational potential, the Schr\"odinger equation and the exact
solutions of the energy levels of the gravitational collapse. 
By using the concept of \emph{BH effective state} \cite{key-1}, 
the quantum gravitational potential, the mass spectrum and the energy spectrum for the extremal
RNBH have been found. 
From the mass spectrum, the area spectrum has
been derived too, and it is consistent with the previous
result by Bekenstein \cite{key-3}.

The stability of these solutions were described with the Majorana
approach (see \cite{key-19} for details) to the Archaic Universe scenario \cite{key-18}, finding
the existence of oscillatory regimes or exponential damping 
for the evolution of a small perturbation from a stable state.

Finally, one takes the chance to recall that in a series of interesting papers \cite{key-20,key-21,key-22}, the authors wrote down the Schr\"odinger equation for a collapsing object and showed by explicit calculations that quantum mechanics is perhaps able to remove the singularity at the BH center (in various space-time slicings); this is consistent with our analysis. Moreover, they also proved (among the other things) that the wave function of the collapsing object is non-singular at the center even when the radius of the collapsing object (classically) reaches zero. In \cite{key-22}, they considered charged BHs. With regards to the area quantization, another interesting approach, based on graph theory, can be found in \cite{key-23}. Here, the Bekenstein-Hawking area entropy accompanied with a proper logarithmic term (subleading correction) is obtained, and the size of the unit horizon area is fixed. Curiously, Davidson also found a hydrogen-like spectrum in a totally different context \cite{key-24}. 

Other interesting approaches can be found in \cite{Vaz1, Vaz2, Makela, Das}.

\section{Acknowledgments}
F.\,T. gratefully acknowledges ZKM and Peter Weibel for the financial support. The authors thank the referee for his/her positive comments on the manuscript.

\end{document}